\documentclass[11pt]{article}
\usepackage{psfig}

\def\be {\begin{equation}}
\def\ee {\end{equation}}
\def\bea {\begin{eqnarray}}
\def\eea {\end{eqnarray}}
\def\n {\nonumber}
\def\bc {\begin{center}}
\def\ec {\end{center}}

\def \mf{{m_{1}}^{2}}
\def \ms{{m_{2}}^{2}}
\def \m3{\left |{m_{3}}\right |}
\def \mths{\left | m_{3}^{2}\right |}
\def \HF{\left |{H_{1}}\right |}
\def \H2{\left |{H_{2}}\right |}
\def \gpsq{{g'}^{2}}
\def \g2sq{{g_{2}}^{2}}
\def \phialpha{{\phi}_{\alpha}}

\def \MHU{m_{H_1}^2}
\def \MHD{m_{H_2}^2}
\def \MHF{m_{1/2}}
\def \MSX{m_{16}}
\def \MTN{m_{10}} 

\def \MGL{m_{\tilde g}}
\def \MSQ{m_{\tilde q}}
\def\lapp{\mathrel{\rlap{\raise.5ex\hbox{$<$}}
                    {\lower.5ex\hbox{$\sim$}}}}
\def\gapp{\mathrel{\rlap{\raise.5ex\hbox{$>$}}
                    {\lower.5ex\hbox{$\sim$}}}}

\begin{document}
\title{
\bf Yukawa Unification and Unstable Minima of the Supersymmetric
Scalar Potential}
\author{ {\large\sl Amitava Datta}
\thanks{Electronic address: adatta@juphys.ernet.in}
\thanks{On leave of absence from Jadavpur University.}\\
Department of Physics, Visva-Bharati, Santiniketan - 731 235, India
\and
{\large\sl Anirban Kundu} 
\thanks{Electronic address: akundu@juphys.ernet.in}
\ \ and {\large\sl Abhijit Samanta} 
\thanks{Electronic address: abhijit@juphys.ernet.in}\\
Department of Physics, Jadavpur University, Calcutta - 700 032, India 
}

\date{\today\\ hep-ph/0007148}
\maketitle

\begin{abstract}
Motivated by the possibilities of $b-\tau$ or $t-b-\tau$ Yukawa unification  
in the supersymmetric Grand Unified Theories, we consider the dangerous 
directions of the supersymmetric potential for large values of $\tan\beta 
~(\gapp 30)$, in two versions of the minimal supergravity model with and 
without common soft breaking scalar masses at the GUT scale, where the 
potential may become unbounded from below.  We find that for the common
trilinear coupling $A_0 \lapp 0$ the requirement of $b-\tau$ unification 
in conjunction with the stability condition on the potential yields highly 
restrictive sparticle spectra with upper, and in many cases, lower bounds 
stronger than the available experimental lower bounds, on the soft SUSY 
breaking common scalar mass and the common gaugino mass. Over a significant 
region of the parameter space, the model becomes even more restrictive if 
the common sfermion soft mass is different from the soft mass for the Higgs 
sector.  We also find that the bulk of this restricted parameter space can
be probed at the LHC. In models with $t-b-\tau$ Yukawa unification, $A_0
\leq 0$ is ruled out from potential constraints. 

\end{abstract}

%
%
\newpage

\pagestyle{plain}   \pagenumbering{arabic}
\setcounter{footnote}{0}
\renewcommand{\thefootnote}{\arabic{footnote}}

\section{Introduction}

If we have to go beyond the Standard Model (SM), for which there are ample 
motivations, the most popular choice seems to be supersymmetry (SUSY) \cite
{SUSY-review}. With a 
plethora of new degrees of freedom, it is necessary to constrain them,
in addition to direct searches at the colliders, in 
as many ways as possible so that the parameter space for the SUSY particles 
may be narrowed down. One of the most useful ways to put such constraints 
is to consider the dangerous directions of the scalar potential where the 
potential may be unbounded from below (UFB) or develops a charge and/or color
breaking (CCB) minima \cite{oldufb}. This may happen 
since one now has charged and colored 
scalar fields in the spectrum, and the possible existence of such a direction
would make the standard vacuum unstable. Different directions are chosen by
giving vacuum expectation value (VEV) to one or more scalar fields, while 
keeping the VEVs of the other scalars to zero.  

Such constraints, in fact, are very powerful. This may be realized from the
fact that the allowed parameter space (APS) for SUSY models
is practically unrestricted  as one goes for larger and larger values of the 
soft SUSY breaking parameters ({\em e.g.}, the universal
scalar and gaugino masses, the trilinear coupling, etc.)\footnote{
Apart from the fact that they should not be more than a few TeV if we have
to have an acceptable solution to the hierarchy problem, there is 
no hint from the theory about their actual values.} beyond the kinematic
limit of the current high energy colliders.
On the otherhand the UFB and CCB constraints quite often acquire 
greater eliminating power to rule out significant parts of
 such  regions beyond the 
striking range  of the current experiments. Thus, there is an
intricate balance between such `potential constraints' and the expanding
SUSY APS. For some values of the free parameters, the UFB and
CCB conditions are very sharp and disallow most of the parameter space 
that is otherwise allowed; for some other values, they lose their 
constraining power.

In a very interesting  paper which revived interest in 
UFB and CCB constraints, Casas {\em et al} \cite{casas} investigated such
constraints on SUSY models. Though their 
formulae  are fairly model-independent, they have carried out
the numerical analysis  for
 moderate values  of $\tan\beta$ (the ratio of the vacuum expectation
values (VEV) of the two Higgs fields) only, when one can ignore the
effects of b and $\tau$ Yukawa couplings in the relevant
renormalization group equations (RGE's).
Further they have used the standard minimal supergravity
(MSUGRA) assumption of universal soft scalar mass $m_0$ and universal
gaugino mass $m_{1/2}$ at the GUT scale $M_G$, referred to hereafter 
as the `conventional scenario', to determine the sparticle 
spectrum. Their main result was that within the framework of MSUGRA, a certain
UFB constraint known as UFB-3 with VEV given in the direction of the 
slepton field  puts the tightest bound on the SUSY parameter
space that they considered (see eq. (93) of \cite{casas} and the discussions 
that follows).

The purpose of this work is to extend  and complement the work of \cite{casas}
by analyzing the effectiveness of the UFB constraints for large $\tan\beta$,
This we have done in two models: (i) the conventional scenario and (ii)
a modified version of MSUGRA within the frame of a SO(10) GUT
where the sfermion soft masses $m_{16}$ are universal at the GUT 
scale, but the
Higgs soft masses $m_{10}$ are diferent from them (this we will call  the
`nonuniversal scenario'). In course of this work we have realized that
in contrast to the low $\tan\beta$ scenario, the UFB-3 constraint with
squarks (eq.\ 31 of \cite{casas}) may become stronger under certain
circumstances, and over a large part of the parameter space the constraint
known as UFB-1 (see eq. (\ref{ufbone})) serves as the chief restrictor 
of the APS.

It is well known that there are quite 
a few motivations for going beyond small and intermediate values of 
$\tan\beta$ in the context of Grand
Unified Theories (GUT). If one assumes the GUT group SO(10) breaking directly
to the SM gauge group SU(3)$\times$SU(2)$\times$U(1), and
a minimal Higgs field content (only one {\bf 10} containing
both the light Higgs doublets required in MSSM),  the top, bottom and $\tau$
Yukawa couplings must unify to a definite GUT scale value at the scale where 
SO(10) breaks \cite{sotenbr}. Within the framework of GUTs partial $b-\tau$
Yukawa unification is also an attractive possibility \cite{b-tau,partial}. 
In an SO(10) model, 
even if one assumes more than one {\bf 10}-plet of Higgs fields,
$\tau$ and bottom Yukawa couplings must unify, but the top Yukawa may not
unify with them at the GUT scale $M_G$. 

It can be shown that $\tan\beta$ must lie in the range
45-52 for $t-b-\tau$ Yukawa unification (for $m_t=175$ GeV)
and in the range 30-50  for only $b-\tau$ unification. We do not consider
the possibility $\tan\beta\leq 2$ since such low values of $\tan\beta$ 
are now under pressure due to the lower bound on the lightest Higgs boson
mass from LEP \cite{barenboim}.  This justifies the enthusiasm that has been
generated regarding the  phenomenology of large $\tan\beta$ scenario 
\cite{baer,denegri}.
    
To motivate the nonuniversal scenario under consideration, let us note that 
from a SUGRA point of view, it is natural to choose the scale at which SUSY 
breaks in the vicinity of the Planck scale $M_P\approx 2.4\times 10^{18}$ 
GeV. At this scale, one may have  truly universal soft  masses for all 
scalars; however, the running of the scalar masses between $M_P$ and $M_G$ 
can lead to a nondegeneracy at $M_G$ \cite{running}. Within the framework 
of an SO(10) GUT, the first two sfermion generations will still be degenerate, 
as they live in the same representation of SO(10) and have negligible 
Yukawa couplings. The Higgs fields live in a different representation, and 
couple to other heavy GUT fields to generate the doublet-triplet splitting; 
so their masses can change
significantly. The third generation sfermions may have a large Yukawa
coupling and hence may be nondegenerate from the first two generation of
sfermions, though this effect has not been taken into account in our
discussion for simplicity. Only the Higgs mass parameter 
($m_{10}$) at $M_G$ is assumed to be 
different from the common soft sfermion mass ($m_{16}$) at that scale,
and both are treated as free parameters.

In addition to restricting the values of $\tan\beta$, 
the requirement of Yukawa unification 
eliminates a significant region of  the otherwise large APS  of MSSM  
quite effectively. For example, this unification occurs within 
a rather limited region of the $m_{16}-m_{1/2}$ plane 
for certain generic choices of the common trilinear soft 
breaking parameter $A_0$. This dependence arises largely through the 
radiative corrections to the running bottom quark mass \cite{pierce}
which in turn controls the bottom quark Yukawa coupling $\lambda_b$ at
low energies. 
The UFB-1 and the UFB-3 conditions further eliminate a significant
region from this already restricted APS, which is one of the 
main results of this paper. Throughout the paper we ignore the possibility
that the nonrenormalizable effective operators may stabilise the 
potential \cite{nonrenorm}.

The APS obtained by requiring Yukawa unification only is
quite sensitive on the choice
of $A_0$. For example, in the conventional scenario with  $b-\tau$ 
unification, the APS increases quite a bit  for  large 
negative values of $A_0$. It is precisely these values of $A_0$ which makes 
the potential more vulnerable to the UFB conditions and many of the 
additional points allowed by choosing $A_0$ appropriately are
eliminated by the UFB conditions, as will be demonstrated 
in a later section. Thus, there is a nice complementary behaviour: for
large negative $A_0$, the Yukawa unification criterion is a weak condition
but UFB conditions are very strong, while for small negative values of
$A_0$ the roles are reversed. For positive $A_0$, none of these criteria 
are sufficiently strong. 

Following the same procedure, significant regions of the parameter space 
can be eliminated for models with $t-b-\tau$ Yukawa unification. In
particular, we find that $A_0\leq 0$ is completely ruled out. 

The effectiveness of Yukawa unification as a restrictor of the APS
also diminishes, as expected, as the accuracy with
which we require the unification to hold good is relaxed.
There are several reasons why the unification may not be exact. 
First, there may be threshold corrections \cite{threshold}, both at the 
SUSY breaking scale (due to nondegeneracy of the sparticles) and at $M_G$, 
of which no exact estimates exist. Secondly, 
we have used two-loop RGE's for gauge
couplings as well as Yukawa couplings and one loop RGEs for the soft 
breaking parameters, but higher order loop corrections may be important 
at a few percent level at higher energy scales. 
Finally the success of the unification program is also dependent 
on the choice of $\alpha_s(M_Z)$  
which is not known as precisely as $\alpha_1$ or $\alpha_2$.
To circumvent such drawbacks, one relaxes the Yukawa unification condition 
to a finite amount (5\%, 10\% or 20\%) which should indirectly 
take care of these possible caveats.  It is interesing to note that 
quite often the  UFB constraints  rule out  subtantial parts 
of the extended APS. 

Some of the ``potential'' constraints analyzed here were also discussed by
Rattazzi and Sarid \cite{sotenbr}. However, they considered the RG improved
tree-level potentials only and included the possibility of stabilizing the
potential through nonrenormalizable effective operators. Moreover the 
potent UFB-3 constraint was not available at the time of their analysis. 
Finally a systematic analysis of the APS in the $\MHF$-$\MSX$ plane, which
is very relevant for physics studies at the Large Hadronic Collider
(LHC), was not presented.

When the SO(10) symmetry breaks down to the SM symmetry, there may be a 
nonzero D-term, which causes the mass splitting between sfermions in {\bf 5}
and {$\overline{\bf 10}$} of SO(10) \cite{d-term}. 
Recently, a number of authors addressed to the phenomenology of the
SO(10) D-terms \cite{pheno-d}.  In this paper, we take the D-term to
be zero for simplicity; with a nonzero D-term, one gets a wider variety
if constraints which will be discussed in a subsequent paper \cite{abhijit}.

It is well-known that there is a basic conflict between $b\rightarrow s\gamma$
and $t-b-\tau$ Yukawa unification. The latter works best for $\mu<0$ and
large $\tan\beta$, while at the same time this region of the parameter
space tends to give unacceptable contributions to the former \cite{borzumati}.
However, in view of the uncertainties in the long-distance corrections and the
possibility of cancellation between various diagrams, we have not included this
constraint in our analysis. 

The plan of the paper is as follows. In the next section, we outline the
various UFB directions of the supersymmetric potential, and discuss our
methodolgy. The next section deals with the results and in the last section,
we summarize and conclude.


\section{UFB Directions of the SUSY potential}

In this section we briefly review the necessary formulae for the UFB directions 
following \cite{casas} and \cite{gamberini}. We closely follow the former
reference in defining the said directions.

The scalar potential of the MSSM is a function of several scalar fields.  An
SU(2)$\times$U(1) breaking minimum of this potential must exist
for preserving the phenomenological successes of the SM. Moreover,
one demands this real minimum $V_{realmin}$ to be deeper than 
the unwanted UFB and CCB minima. These minima are computed by giving VEV to
one or more scalar components at a time; the condition is that at no point
in such dangerous directions the potential should be deeper than $V_{realmin}$.
The resulting constraints on the field space are of much importance  
as they can restrict the soft SUSY breaking parameters,
and hence the sparticle masses and couplings \cite{oldufb}.  
Let us see how these dangerous field directions arise.

The tree level scalar potential in the  MSSM can be written as the sum of
the D-term, the F-term and the soft mass term: 
\be
V_0=V_F+V_D+V_{soft}
\ee
where
\bea
V_F&=&\sum_{\alpha} \left |\frac{\partial W}{\partial \phi_\alpha} \right |^2,
\n\\
V_D&=&\frac{1}{2} \sum_a{g_a^2}\left( \sum_\alpha{\phi_\alpha^\dag}
T^{a}\phialpha \right) ^2,\n\\ 
V_{soft}&=&\sum_\alpha m_{\phi_\alpha}^{2} \left|
\phi_\alpha\right|^{2} + \sum_i
\{ A_{u_i}\lambda_{u_i}Q_i H_2u_i+A_{d_i}\lambda_
{d_i}Q_iH_1d_i+A_{e_i}\lambda_{ei}L_iH_1e_i+h.c.\}\n\\
&{ }&+(B\mu H_1H_2+h.c)
    \label{soft}
\eea
and the superpotential $W$ is defined as
\be
W= \sum_i \{\lambda_{u_{i}}Q_{i}H_{2}u_{i}+\lambda_
{d_{i}}Q_{i}H_{1}d_{i}+\lambda_{e{i}}L_{i}H_{1}e_{i}\}
+\mu H_{1}H_{2}.
    \label{superpot}
\ee
Here,  $\phi_{\alpha}$ are the generic scalar fields,
$T_{a}$ and  $g_{a}$ are the gauge group generators and the gauge couplings
respectively, and $\lambda$'s are the respective Yukawa couplings.
$A_i$, $B$ and $m_i$ are the soft SUSY breaking parameters, and $\mu$ is
the Higgsino mass term. In eq. (\ref{superpot})
$Q_i$ and $L_i$ stand for SU(2) doublet squark and slepton
superfields while $u,d$  and $e$ are the corresponding singlet superfields.
The generation index $i$ runs from 1 to 3.

The neutral part of the Higgs potential in the MSSM is given by
\be
V_{Higgs}=m_{1}^{2}\left | H_{1}^{2}\right | + m_{2}^{2}\left|H_{2}^{2}
\right|-2 \left | m_{3}^{2}\right|\left | H_{1}\right | \left | H_{2}\right |
+\frac{1}{8}  
(g'^{2}+g_{2}^{2})(\left|H_{2}\right|^{2}-\left|H_{1}\right|^{2})^{2}
\ee
where $m_1^2=m_{H_1}^2+\mu^2$, $m_2^2=m_{H_2}^2+\mu^2$, and
$m_3^2=-\mu B$, with $m_{H_1}$ and $m_{H_2}$ being the mass terms of the
two doublets. The standard GUT normalization is used for the gauge couplings:
$g_3=g_2=g_1=\sqrt{5/3}g'$ at $M_G$.
Minimization of this tree-level potential yields
\be
V_{realmin}=-\frac{\{[{(\mf+\ms )}^2-4{\m3}^4]^{1/2}-
\mf+\ms\}^2}{2(\gpsq+\g2sq)}.
\ee
At any scale $Q$, there is a significant radiative correction to this 
potential. Including the one-loop corrections, the potential becomes
\be
V=V_0+\Delta V_{1}
\ee
where
\be
\Delta V_{1}=\sum_\alpha \frac{n_\alpha}{64 \pi} M_{\alpha}^4\left[\ln\frac
{M_\alpha^2}{Q^2}-\frac{3}{2}\right],
   \label{oneloop}
\ee 
with $n_{\alpha}=(-1)^{2s_{\alpha}}( 2s_{\alpha}+1)$, 
$s_{\alpha}$ being the spin of the corresponding field.
One ensures the minima of $H_1$ and $H_2$ at
$\HF=v_{1}$ and $ \H2=v_{2}$ with $M_W^2=\frac{1}{2}g^2 (v_1^2+v_2^2)$.

As was pointed out, the constraints on the APS arise from directions 
in the field space along which the potential 
becomes lower than $V_{realmin}$ (and may become unbounded from below).  
However, the minimization  of the full potential $V$
is rather cumbersome. On the otherhand, just minimizing the tree-level 
potential at the weak scale neglecting $\Delta V_1$ can lead to  
quite erroneous conclusions about the minimum point in the field space 
as was shown by \cite{gamberini}. As a compromise, one evaluates $V$ at
a judiciously chosen scale $\hat Q$ where the one-loop correction is
minimum. As is evident from eq. (\ref{oneloop}), this scale should be 
about the typical SUSY mass scale $M_S$ so that the large logarithmic terms 
tend to vanish.

The dangerous directions are selected in such a way 
that the positive definite F-terms
vanish and the D-terms either cancel each other or their magnitudes 
can be kept under control. There are several other guidelines as discussed 
in \cite{casas}.  Using these conditions one can get the following 
UFB potentials.

\vspace*{2mm}

{\bf UFB-1}: The condition 
\be
\mf+\ms \geq 2 \mths, 
    \label{ufbone}
\ee
which is known as UFB-1, must be satisfied at any scale $\hat{Q}>M_S$,  
and particularly at the unification scale $\hat{Q}=M_G$, to have a realistic
minimum of the scalar potential. From eq. (\ref{ufbone}),
small $\MHU$ (and $\MHD$) makes the UFB-1 condition severely restrictive. 
This may be the case for large $\tan\beta$ and large negative values of $A_0$.
The variations of $m_{H_1}^2$ and $m_{H_2}^2$ in the conventional scenario
with respect to the common trilinear coupling 
$A_0$ for  $\tan\beta=30$ and 45, corresponding to $b-\tau$ and $t-b-\tau$ 
Yukawa unification respectively,  are illustrated in fig.\ 1.
From the figure we find that negative values of $A_0$ drive
$m_{H_2}^2$ to large negative values in both the cases. For $\MHU$
the effect is prominent for large $\tan\beta$, while for $\tan\beta=30$, 
$\MHU$ remains positive for most of the range of $A_0$ that we have studied.
The plot of $m_H^2$ vs $A_0$ is also given for different values of 
$m_{1/2}$ and $m_{16}$ for $\tan\beta=45$ in figures 2 and 3 respectively.
From these plots it is clear that for large $m_{1/2}$ and/or $m_{16}$, and
large negative $A_0$, $m_{H_1}^2$ and $m_{H_2}^2$ decrease significantly, 
so that these values of $m_{1/2}$ and $m_{16}$ become vulnerable to UFB-1.
This is the reason why in the large $\tan\beta$ case the UFB-1
constraint plays a very significant role in restricting the APS.

\begin{figure}[htb]
\centerline{
\psfig{file=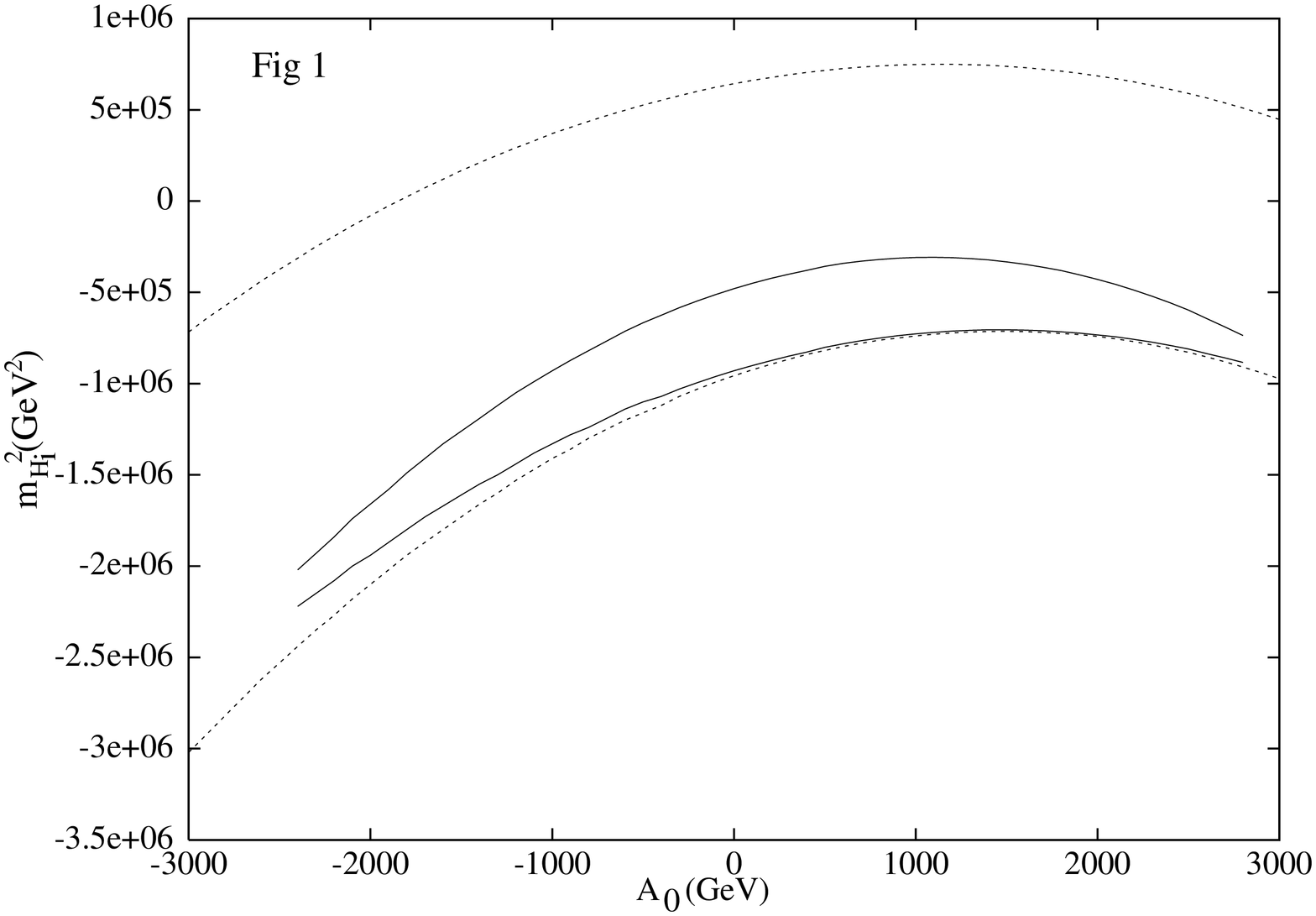,width=12cm,angle=270}
}
\caption{\sl{The variation of the Higgs mass parameters $m_{H_1}^2$ and
            $m_{H_2}^2$ with the trilinear coupling $A_0$. 
            The solid (dotted) lines are for $\tan\beta=45 (30)$.
            The top two lines are for $m_{H_1}^2$ while the lower pair 
            is for $m_{H_2}^2$. We have used $m_{16}=m_{10}=m_{1/2}=1$ TeV. 
            }}
    \label{fig:figure1}
\vspace*{5mm}
\centerline{
\psfig{file=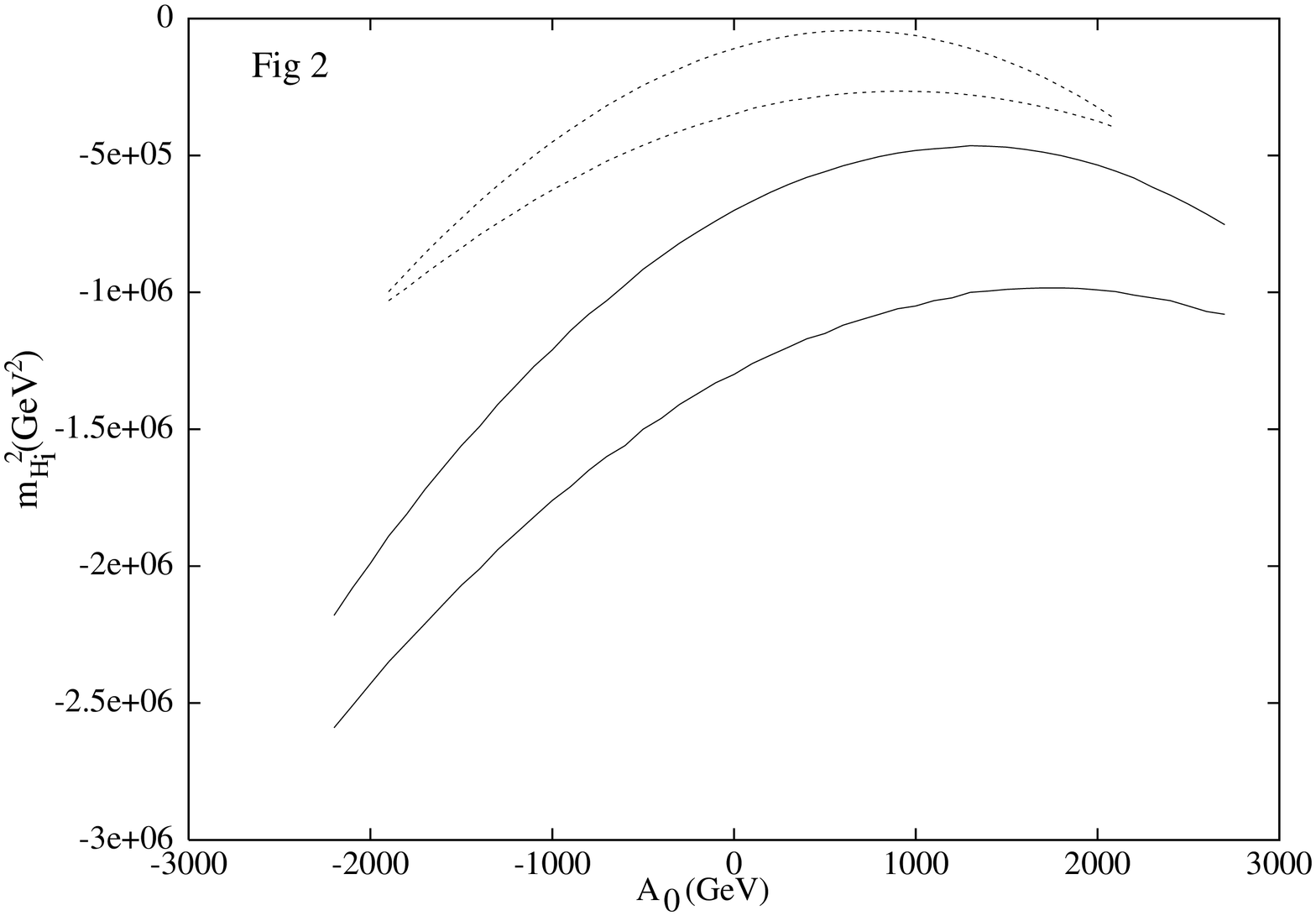,width=12cm,angle=270}
}
\caption{\sl{The same variation as shown in figure 1, with  
             two different values of $m_{1/2}$. The dotted (solid) pair 
             is for $m_{1/2}= 600 (1200)$ GeV. The upper line in each pair 
             is for $m_{H_1}^2$ and the lower one for $m_{H_2}^2$. 
             $m_{16}=m_{10}=1$ TeV, $\tan\beta=45$.
            }}
      \label{fig:figure2}
\end{figure}
\begin{figure}[htb]
\centerline{
\psfig{file=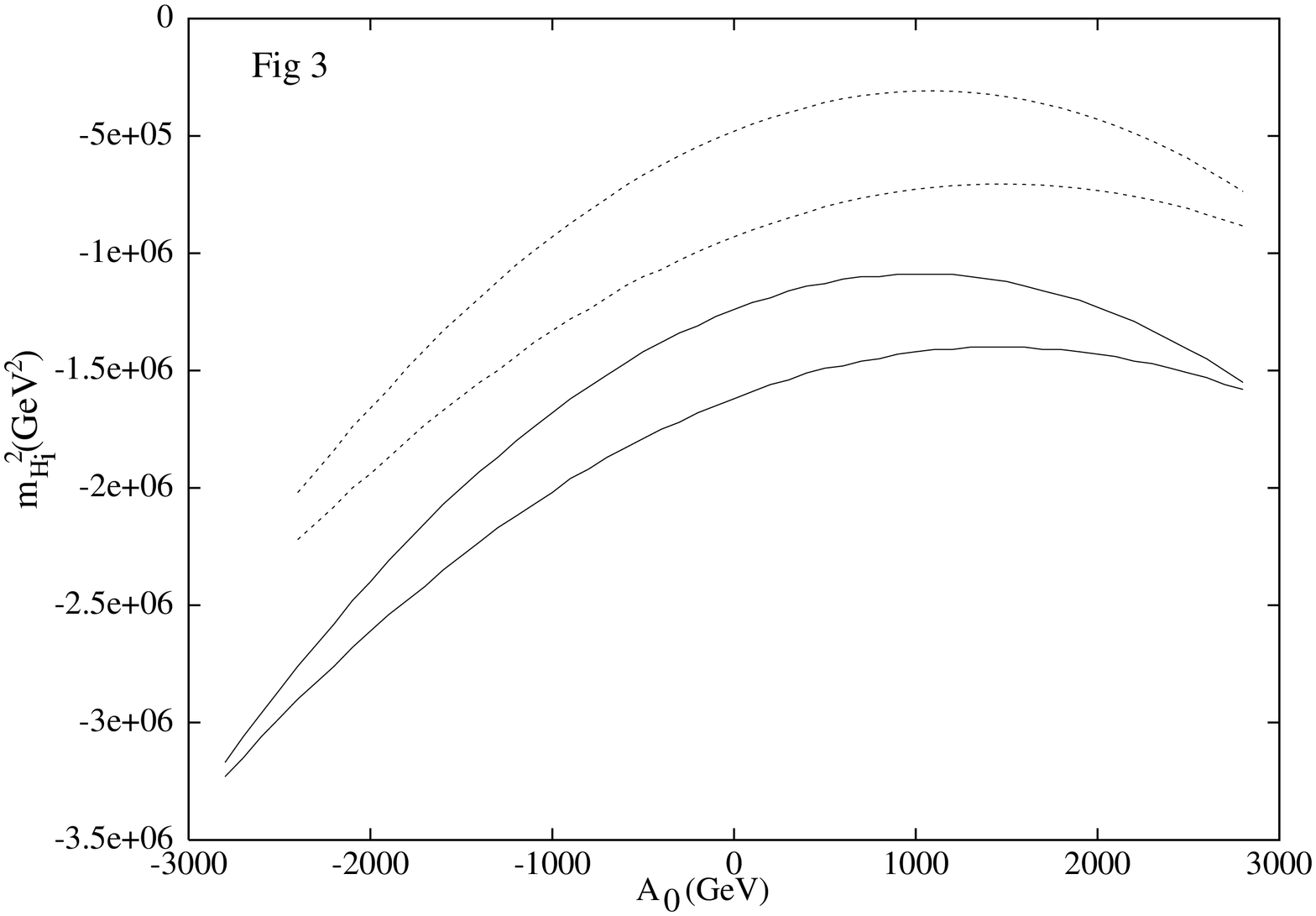,width=12cm,angle=270}
}
\caption{\sl{The same variation as shown in Figure 1, with two different
            values of $m_{16}=m_{10}$.  The dotted (solid) pair is for 
            $m_{16}=1 (1.5)$ TeV.  The upper line in each pair
            is for $m_{H_1}^2$ and the lower one for $m_{H_2}^2$.
            $m_{1/2}=1$ TeV, $\tan\beta=45$.
            }}
    \label{fig:figure3}
\vspace*{5mm}
\centerline{
\psfig{file=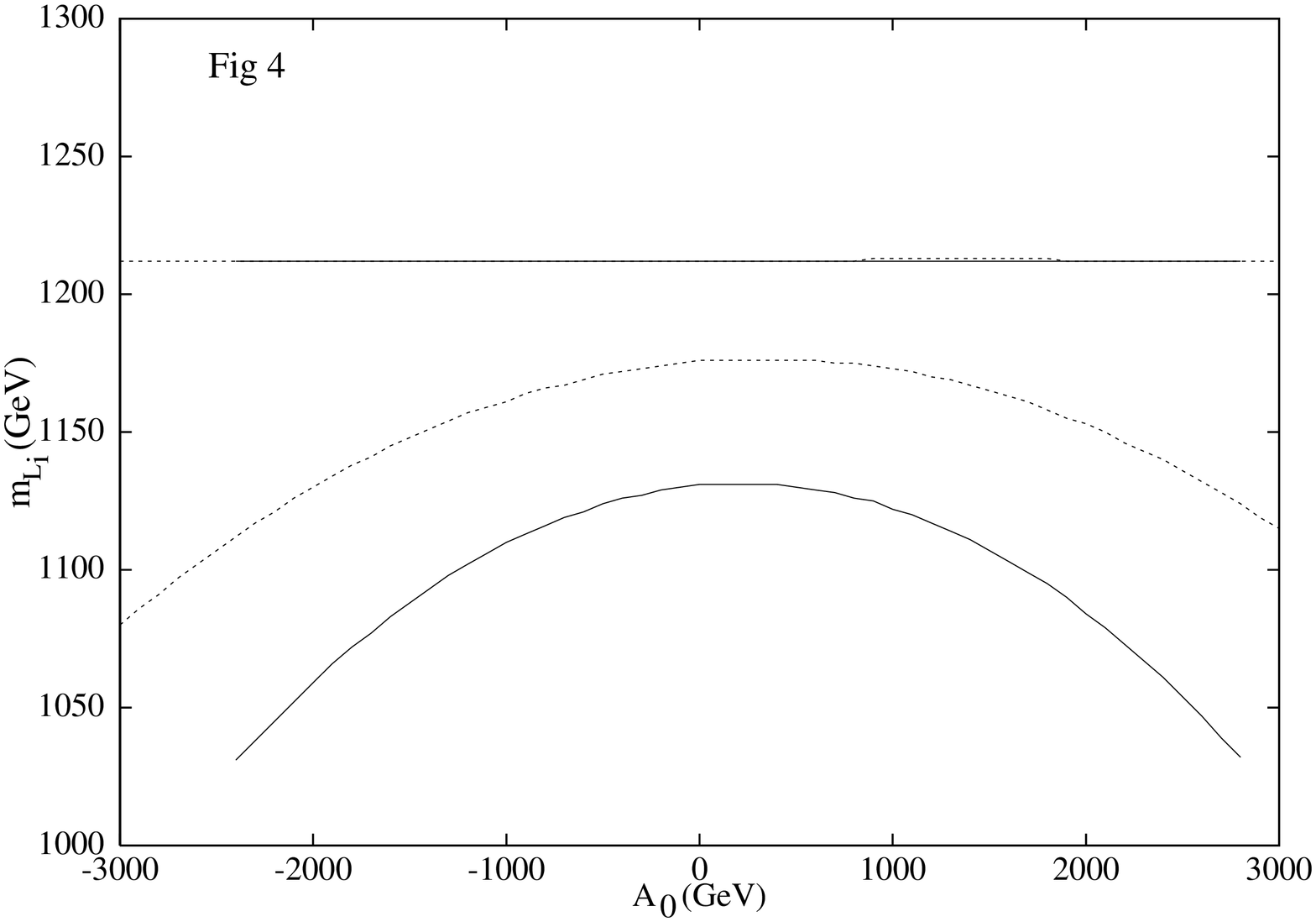,width=12cm,angle=270}
}
\caption{\sl{The variation of left-handed slepton mass parameters with $A_0$.
             The dotted (solid) pair is for $\tan\beta = 30(45)$. In each pair,
             the upper line is for $\tilde{e_L}$ and the lower line for 
             $\tilde{\tau_L}$. Note that the selectron mass is insensitive 
             to the value of $\tan\beta$ and $A_0$.
            }}
      \label{fig:figure4}
\end{figure}
\begin{figure}[htb]
\centerline{
\psfig{file=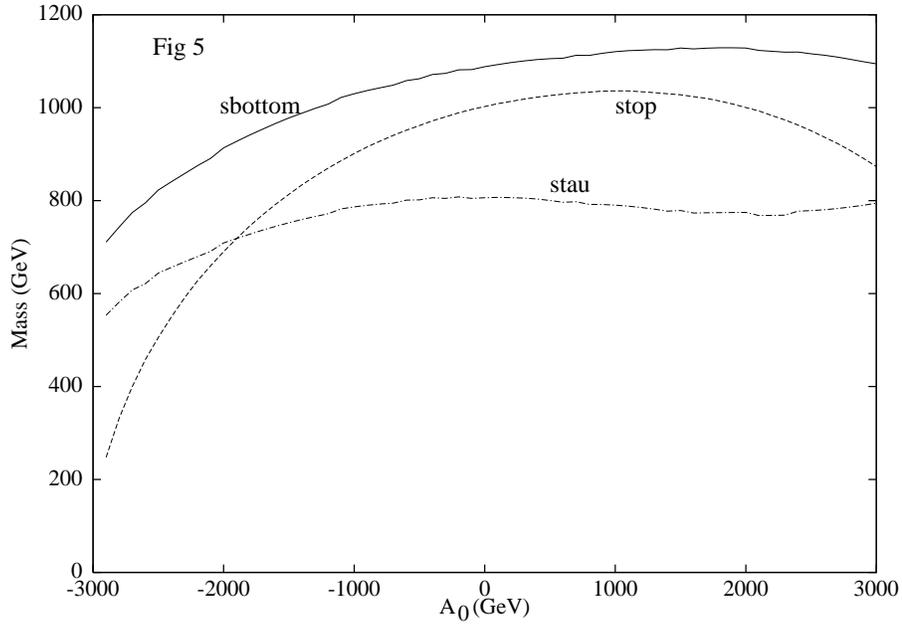,width=12cm,angle=270}
}
\caption{\sl{The variation of masses of the lightest $\tilde\tau$, $\tilde t$
            and $\tilde b$ with $A_0$. We set $m_{16}=m_{10}=1$ TeV
            and $m_{1/2}=500$ GeV. The Yukawa unification condition fixes
            $\tan\beta$. Note that $\tilde t_1$ can be the
            second lightest sparticle for $A_0\lapp -1.8$ TeV.
            }}
    \label{fig:figure5}
\end{figure}

\vspace*{2mm}

{\bf UFB-2}: The doublet slepton (along the sneutrino direction)
and both $H_1$ and $H_2$ are given nonzero VEVs.
For any value of $\H2<M_G$ satisfying 
\be
\left | H_{2} \right |^{2} > \frac{4m_{L_i}^2}{(\gpsq+\g2sq)
[1-\frac{\m3^4} {\mu^4}]},
\ee
and provided that 
\be
\mths<\mu^{2} (= {m_1}^2-{m_{L_i}}^2),
\ee
the UFB-2 potential is given by
\be
V_{UFB-2}=[\ms +m_{L_{i}}^{2}-\frac{\m3^{4}}{\mu^{2}}]\H2^{2}-
\frac{2m_{L_{i}}^{4}}{\gpsq+\g2sq}
\ee
At any momentum scale $\hat Q$, this should be greater than $V_{realmin}$
for a stable configuration:
\be
V_{UFB-2}(Q=\hat{Q})>V_{realmin}(Q=M_{S})
\ee
where $\hat{Q}\sim Max(g_{2}\H2, \lambda_{top}\H2,M_S)$.
However, we find that UFB-2 hardly rules out any further region of the APS
which passes the UFB-1 and UFB-3 constraints, so it is of limited interest 
to us.

\vspace*{2mm}

{\bf UFB-3}: The convention is to choose $H_1=0$ and to cancel the $H_1$
F-term (which is a combination of $H_2$ and $d_{L_i},d_{R_i}$ or
$e_{L_i},e_{R_i}$) with suitable VEVs to $H_2$ and the abovementioned
slepton or squark directions. However, it is economical to give VEVs to the
doublet fields (along $T_3=-1/2$ direction) rather than the singlet fields 
to cancel both SU(2) and U(1) D-terms at the same stroke.
Suppose the sleptons are given VEV; then 
for any values of $\H2<M_G$ satisfying 
\be
\H2>\sqrt{\frac{\mu^{2}}{4\lambda_{e_{j}}}+\frac{2m_{L_{i}}^{2}}{\gpsq+\g2sq}}
- \frac{\left|\mu\right|}{2\lambda_{e_{j}}},
    \label{ufbthree-h}
\ee
the UFB-3 potential is defined as
\be
V_{UFB-3}=[\ms-\mu^{2}+m_{L_{i}}^{2}]\H2^{2}+\frac{\left|\mu\right|}
{\lambda_{e_{j}}}[m_{L_{j}}^{2}
+m_{e_{j}}^{2}+m_{L_{i}}^{2}]\H2-\frac{2{m_{L_{i}}}^{4}}{\gpsq+\g2sq}.
     \label{ufbthree}
\ee
If $\H2$ does not satisfy (\ref{ufbthree-h}), the formula changes to
\be
V_{UFB-3}=[\ms-\mu^{2}]\H2^{2}+\frac{\left|\mu\right|}
{\lambda_{e_{j}}}[m_{L_{j}}^{2}
+m_{e_{j}}^{2}]\H2 + \frac{1}{8} ({\gpsq+\g2sq})\Big[\H2^{2} +
\frac{\left|\mu\right|}{\lambda_{e_{j}}}\H2\Big]^{2}
     \label{ufbthree-alt}
\ee
with $ i\neq j$. Note that we could substitute squarks for sleptons, where
$i=j$ is allowed.  The constraints on the parameter space arise from the 
requirement 
\be
V_{UFB-3}(Q=\hat{Q})>V_{realmin}(Q=M_{S})
\ee
where $\hat{Q}$ is chosen to be $\hat Q\sim Max( g_{2}\left|e\right|,  
g_{2}\H2, \lambda_{top}\H2, g_{2}\left|L_{i}\right|,
M_{S})$ to minimize $\Delta V_1$. The VEVs are not arbitrary; they satisfy
\be
\left| e \right|=\sqrt{\H2 \left|\mu \right|/\lambda_{e{j}}},\ \ 
\left | L_i^2\right |=\left(\H2^2+\left| e \right|^2\right)
 -4{m_{L_i}^2\over (\gpsq+\g2sq)}.
\ee

As can be seen from eq. (\ref{ufbthree}), the region of the parameter space 
where $\MHD={m_2}^2-\mu^2$ is large and negative is very
susceptible to be ruled out by the UFB-3 condition. 
This is because the first term of eq. (\ref{ufbthree})
may become negative in this case. However, the second term in (\ref{ufbthree}),
which is positive definite, may become competitive in certain cases 
({\em e.g.}, for $j=1$, when the Yukawa coupling in the denominator is small), 
which directions one should avoid when looking for the dangerous minima. 

$V_{UFB-3}$ with sleptons was found to be the strongest among
all the UFB and CCB constraints in the low $\tan\beta$ case \cite{casas}.
In order to get the optimum result one has to take the largest $\lambda_{e_{j}}$
in the second term of eq. (\ref{ufbthree}), which leads to the choice 
$e_j = \tilde{\tau}_R$.
Now the restriction $i\neq j$ requires $L_i = \tilde e_L$ or $\tilde \mu_L$
and excludes the choice $\tilde{\tau}_L$. In the low $\tan\beta$ case this 
restriction, however, is of little consequence since all the left sleptons
are degenerate to a very good approximation.

The UFB -3  constraint with squarks  may also be imposed by the following
replacements in (\ref{ufbthree}):
\be
e \rightarrow d, \ \ 
\lambda_{e_j} \rightarrow \lambda_{d_j},\ \ L_j \rightarrow Q_j
    \label{replace}
\ee
(see eq. (31) of \cite{casas}).
Now $i$ may be equal to $j$ and 
$\hat{Q}\sim Max( g_2\left|d\right|,  g_{2}\H2,
\lambda_{top}\H2, g_{2}\left|L_{i}\right|, M_{S})$.

Now the optimum choice is $d_j = \tilde{b}_R$. However, since the choice 
$i=j$ is permitted,  $L_i = \tilde {\tau}_L $ is not excluded. 
At high negative $A_0$ and at high $\tan\beta$, $m_{\tilde{\tau_L}}$ becomes 
smaller than the corresponding mass parameters of the 
first two generations. The variation of left-handed slepton mass parameters
with $A_0$ (for $\tan\beta=45$) is shown in fig.\ 4.
This relatively small $m_{\tilde{\tau_L}}$ at high $\tan\beta$ may make the
alternative choice (\ref{replace}) more restrictive than the UFB-3 
condition with sleptons. This, in fact, has been 
supported by our numerical computations.


\section{Results}

We now  briefly review our methodology for implementing the Yukawa unification
and computing the spectrum which is based on
the computer program ISASUGRA, a part of the ISAJET package.
We use the ISAJET version 7.48 \cite{ISAJET}.

To calculate Yukawa couplings at $\hat Q = M_Z$, we start with the pole masses 
$m_b=4.9$ GeV, $m_\tau =1.784$ GeV and $m_t=175$ GeV.
At $\hat Q = M_Z$ the SUSY loop corrections to $m_b$ and $m_\tau$ is
included using the approximate formulae from ref.\ \cite{pierce}.
For the top quark Yukawa coupling this correction is added at $\hat Q = m_t$.
Starting with the three gauge couplings and the $t,b$ and $\tau$
Yukawa couplings, we evolve them upto the energy scale $M_G$. 
Now the boundary conditions are imposed on the soft breaking parameters
according to the conventional or the nonuniversal scenario,
while trial values for the $\mu$ and $B$ parameters are taken.  
Then all parameters are evolved down to the weak scale $M_Z$. 
The parameters $\mu$ and $B$ are then tentatively fixed at $\hat Q=
\sqrt{m_{\tilde{t_L}}m_{\tilde{t_R}}}$ by the radiative SU(2)$\times$U(1) 
breaking conditions.
Using the particle spectrum so obtained, we compute the radiative corrections 
to the SU(2)$\times$U(1) breaking condition, and hence obtain the corrected 
result for $\mu$ and $B$.
The whole proccess is then repeated iteratively until a stable 
solution within a reasonable tolerance is achieved.  
While running down from $M_G$, the SUSY thresholds
are properly taken care of. The renormalization group (RG)
equations that we use are upto two-loop for both the gauge couplings and
the Yukawa couplings.

The demand of the Yukawa coupling unification at $M_G$
puts an extra constraint on $\tan\beta$. We require unification
within an accuracy of 5\% for $Y_b$ and $Y_{\tau}$ 
and 10\% for $Y_t, Y_b$ and $Y_{\tau}$. The accuracy for the latter 
is relaxed since there are more uncertain factors, {\em e.g.}, the 
choice of the Higgs sector.
We define three variables $r_{b\tau}, r_{tb}$ and $r_{t\tau}$ where generically
$r_{xy}=Max(Y_x/Y_y,Y_y/Y_x)$. To check whether the couplings unify, 
we select only those points in the parameter space where
$Max(r_{b\tau},r_{tb},r_{t\tau})<1.10$ 
(for $t-b-\tau$ unification) and $r_{b\tau}<1.05$ (for $b-\tau$ unification). 
The quark Yukawa couplings depend on $\alpha_s(M_Z)$ which comes out from
the gauge unification conditions to be $0.118$. 
Then we impose the experimental constraints 
$m_{\chi^+}>95$ GeV, $m_h>85.2$ GeV and $m_{\tau_1}>73$ GeV, and
require the lightest neutralino to be the lightest SUSY particle (LSP). 
These constraints filter out the APS on which the potential minima conditions 
UFB-1 and UFB-3 should apply.

Using $\mu,B$, the gauge and the Yukawa couplings at the GUT scale alongwith
the boundary conditions there, we generate the mass spectrum
at any scale $\hat Q$ using the 26 RG equations of the MSSM. 
In fig.\ 5, we show the lightest $\tilde\tau$, $\tilde t$ and $\tilde b$
masses at the weak scale as functions of $A_0$ for the conventional scenario 
with $b-\tau$ Yukawa unification at 
$m_{16}=1$ TeV, $m_{1/2}=500$ GeV (this particular point, for the range of 
$A_0$ shown, is allowed by all constraints that we have considered).
Note that for $A_0\lapp -1.8$ TeV, the lightest stop is the 
next lightest SUSY particle (NLSP), and is perfectly in the accessible
range of the LHC.

We demand the electroweak symmetry to be unbroken at $M_G$.
The Higgs potential is minimized at $\hat Q=\sqrt{m_{\tilde{t_L}}m_{\tilde
{t_R}}}$.
The proper scale for the UFB potential where the one-loop effects are 
minimized, as discussed after eq. (\ref{ufbthree-alt}), is chosen by  
an iterative process within 1\% accuracy. Usually a few iterations are
sufficient. The UFB potential is calculated for different $|H_2|$ values
ranging from zero to $M_G$, using a logarithmic scale.
For each value of $\H2$ we compare the UFB potential with the scalar 
potential of MSSM, and whenever $V_{UFB}<V_{realmin}$,
that particular region in the parameter space is marked as disallowed.

It should be emphasized that if the model is subject to the constraint of 
$b-\tau$ Yukawa unification alone, the allowed region of the $\MHF-\MSX$ 
parameter space increases as $A_0$ becomes more negative. The additional 
regions of the parameter space thus opened up are, however, severely 
restricted by the stability conditions on the potential. As a
result the region allowed by Yukawa unification in conjunction with the 
stability of the potential is restricted to a rather small region even 
for large negative values of $A_0$. This
will be illustrated by the following numerical results.

\begin{figure}[htb]
\centerline{
\psfig{file=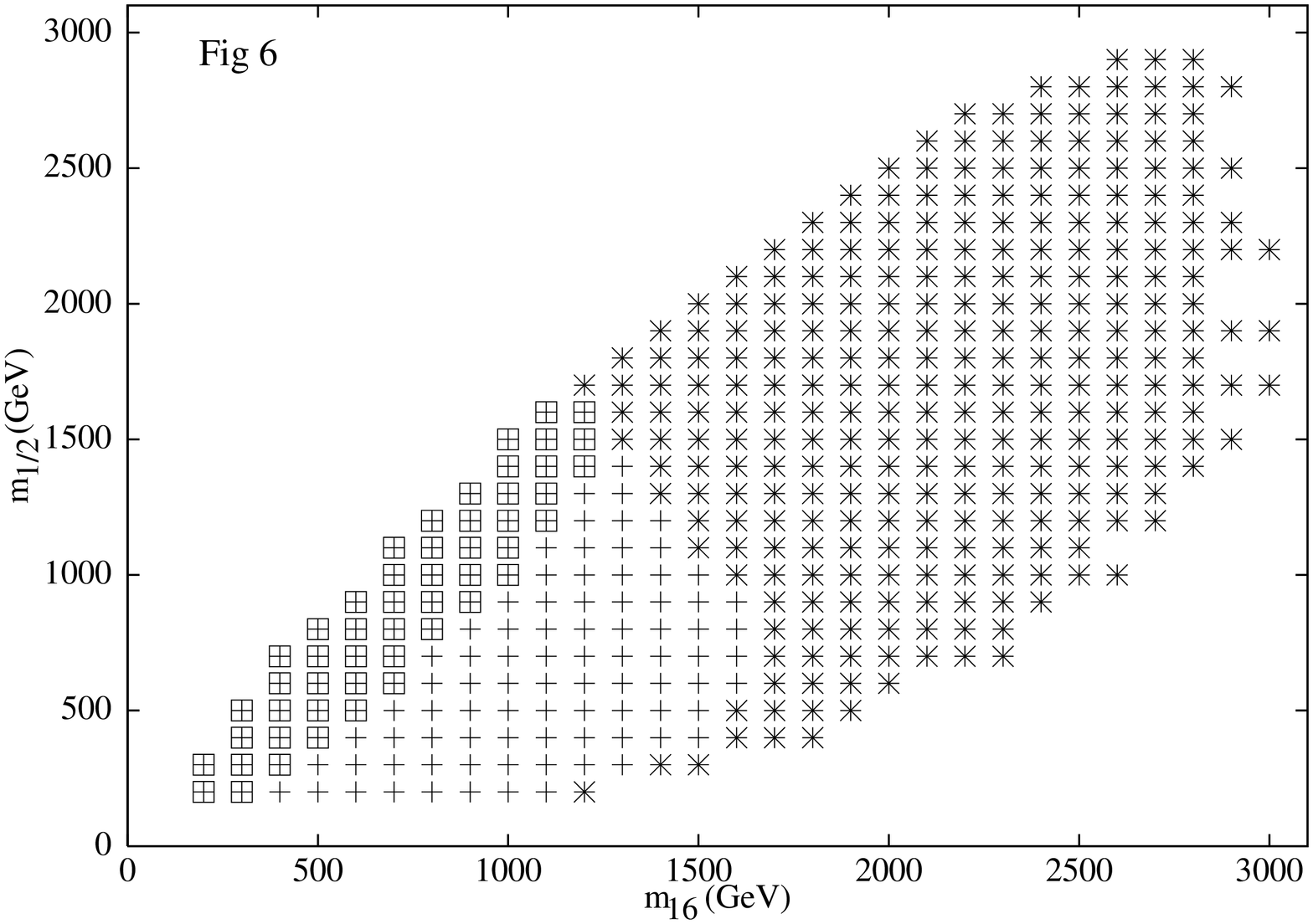,width=12cm,angle=270}
}
\caption{\sl{The allowed parameter space in the conventional scenario with
             $b-\tau$ unification. All the points are allowed by the Yukawa
             unification criterion; the asterisks are ruled out by UFB-1
             and the boxes by UFB-3. We set $A_0=-2m_{16}$.
            }}
    \label{fig:figure6}
\vspace*{5mm}
\centerline{
\psfig{file=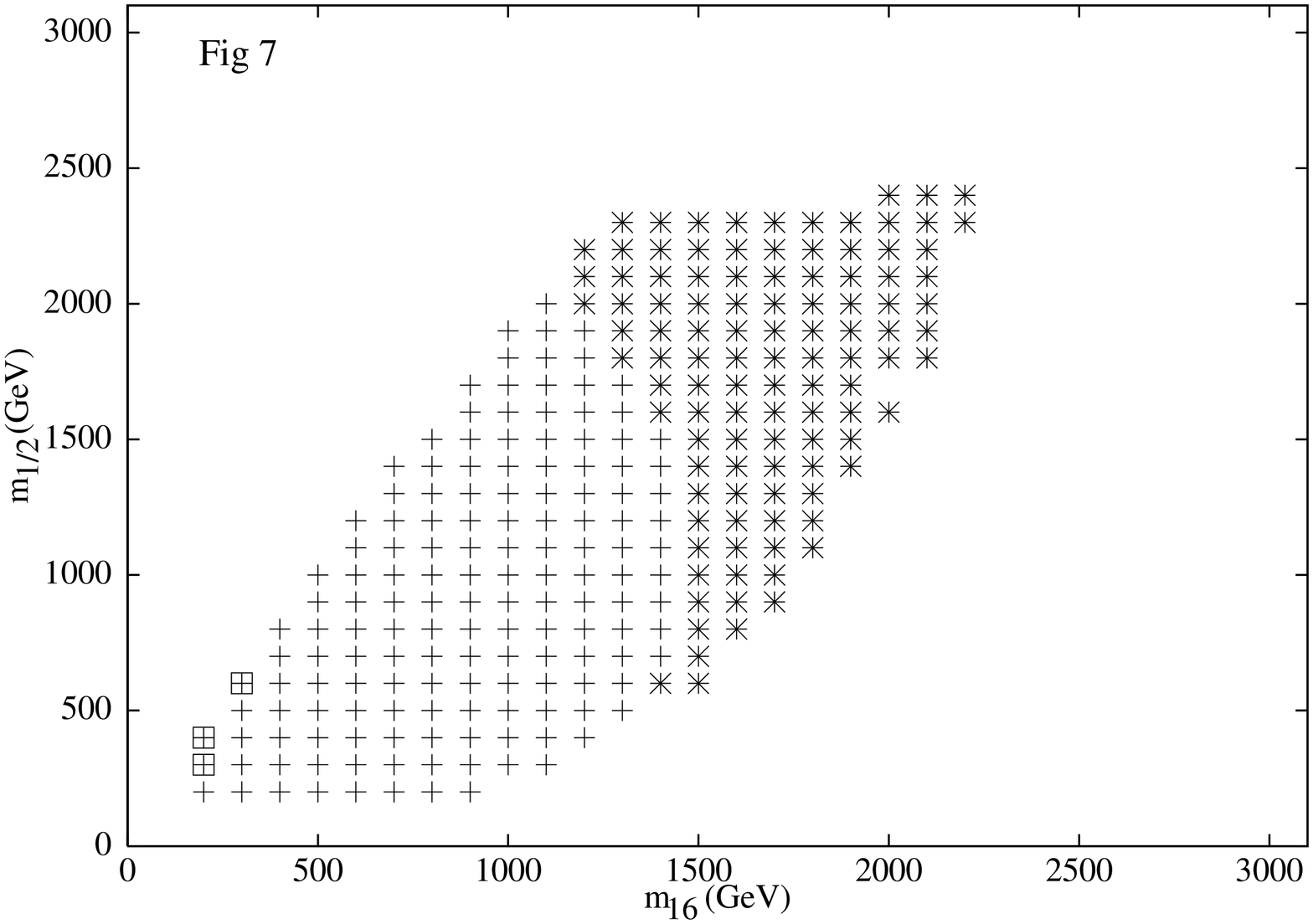,width=12cm,angle=270}
}
\caption{\sl{The same as Fig. 6, with $A_0=-m_{16}$.
           }}
     \label{fig:figure7}
\end{figure}

\begin{figure}[htb]
\centerline{
\psfig{file=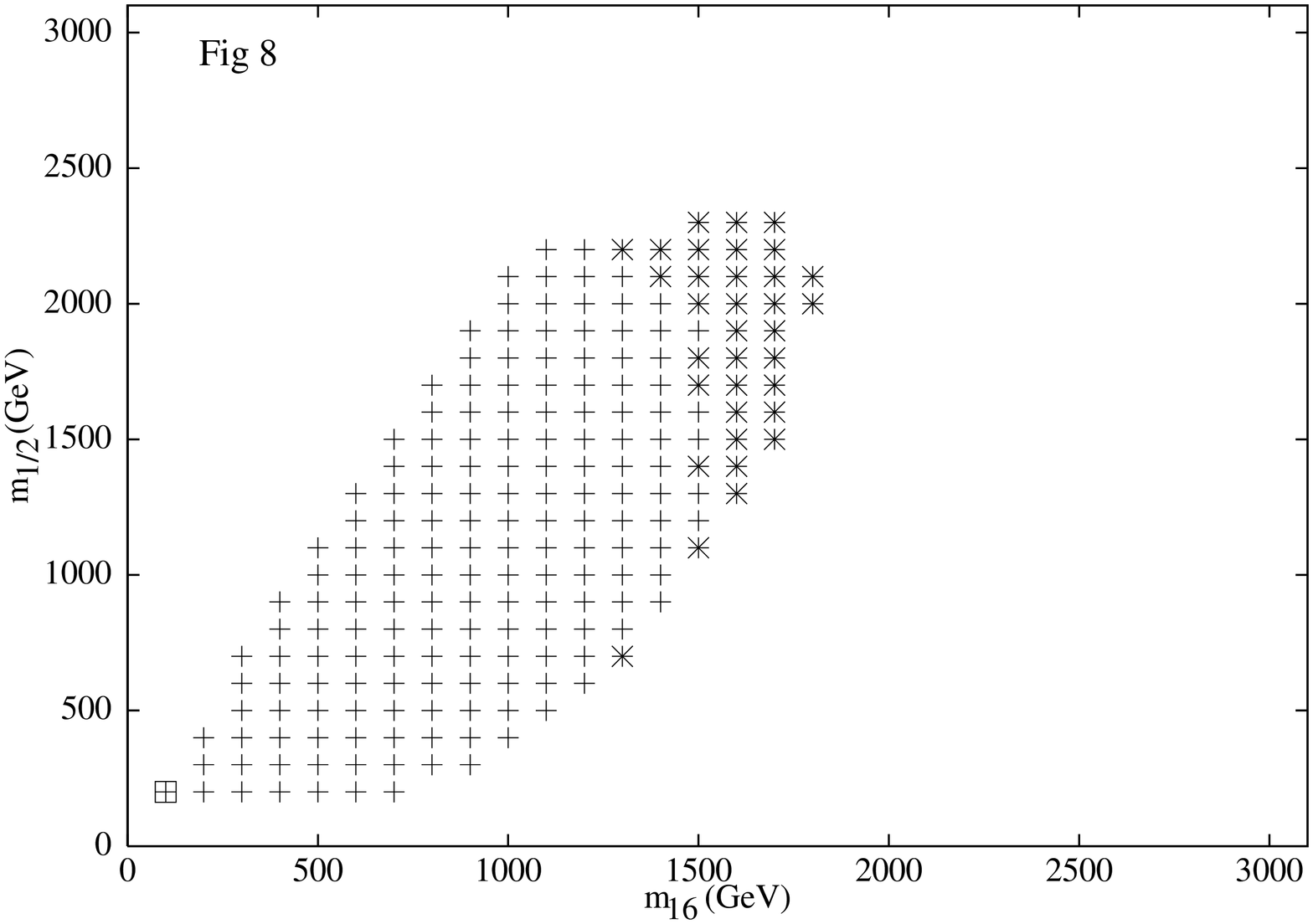,width=12cm,angle=270}
}
\caption{\sl{The same as Fig. 6, with $A_0=0$.
           }}
     \label{fig:figure8}
\end{figure}

\begin{figure}[htb]
\centerline{
\psfig{file=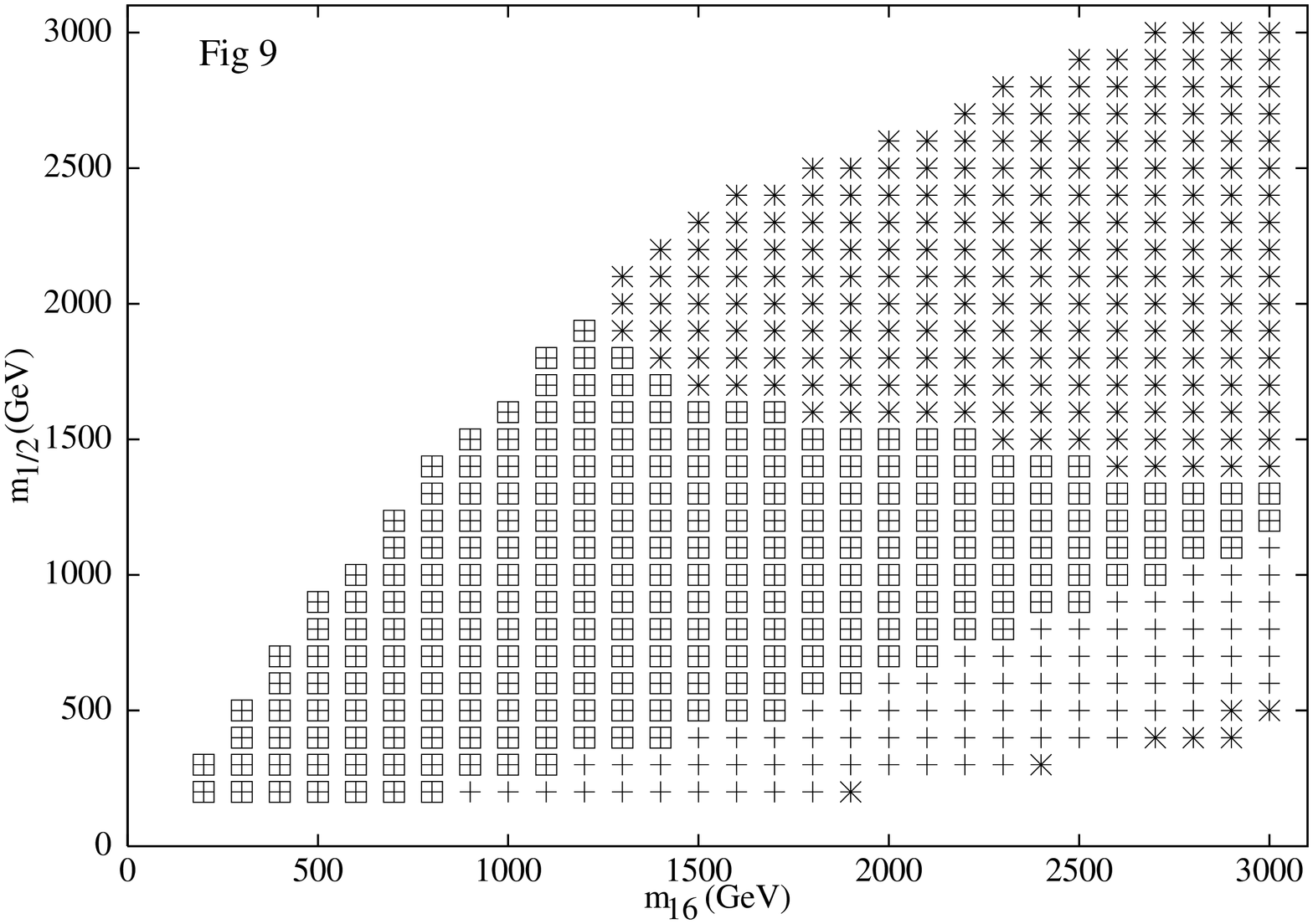,width=12cm,angle=270}
}
\caption{\sl{The allowed parameter space in the nonuniversal scenario with
             $b-\tau$ unification. All the points are allowed by the Yukawa
             unification criterion; the asterisks are ruled out by UFB-1
             and the boxes by UFB-3. We set $A_0=-2m_{16}$ and $m_{10}=
             0.6m_{16}$.
            }}
    \label{fig:figure9}
\vspace*{5mm}
\centerline{
\psfig{file=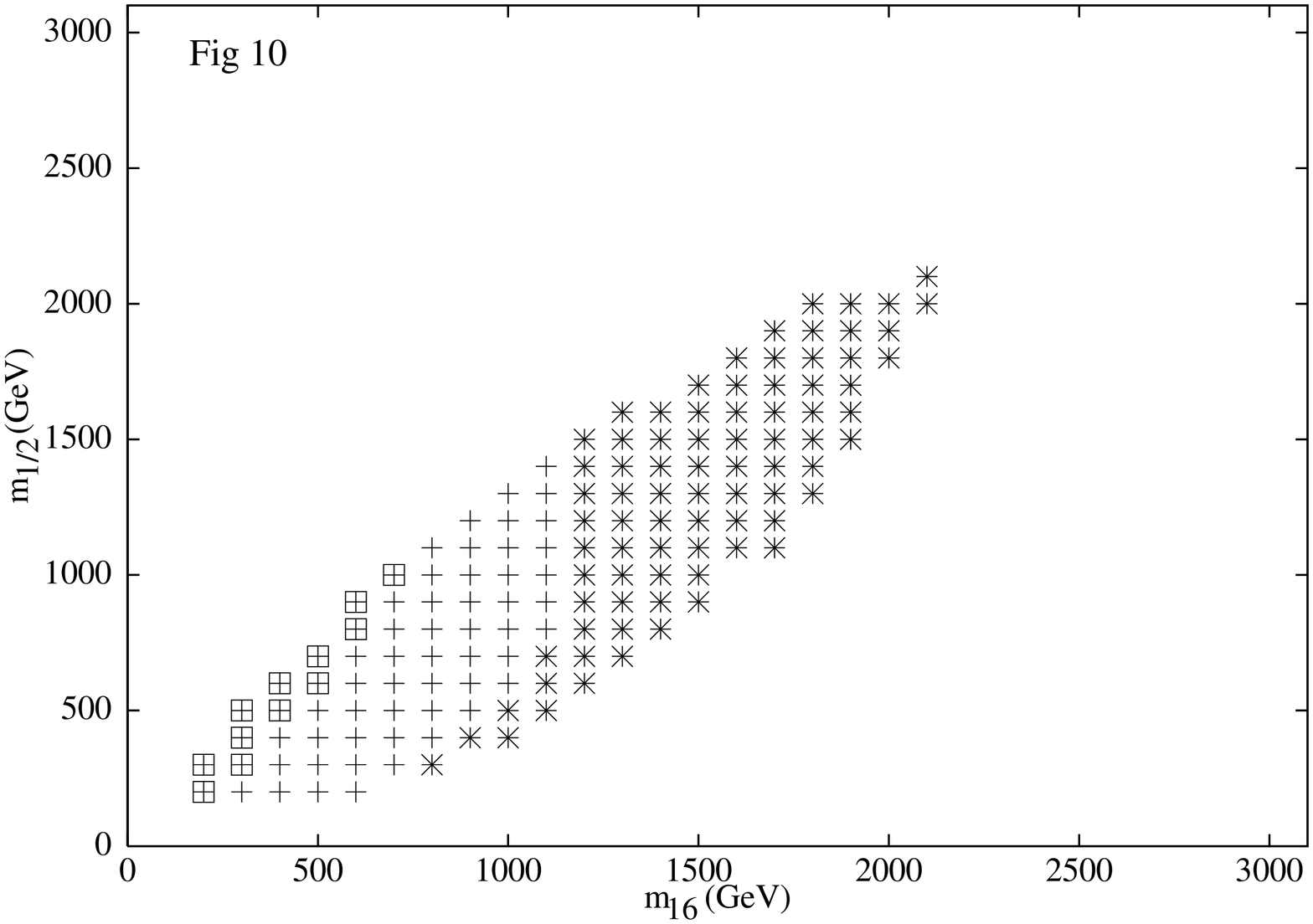,width=12cm,angle=270}
}
\caption{\sl{The same as Fig. 9, with $m_{10}=1.2m_{16}$.
           }}
     \label{fig:figure10}
\end{figure}

\begin{figure}[htb]
\centerline{
\psfig{file=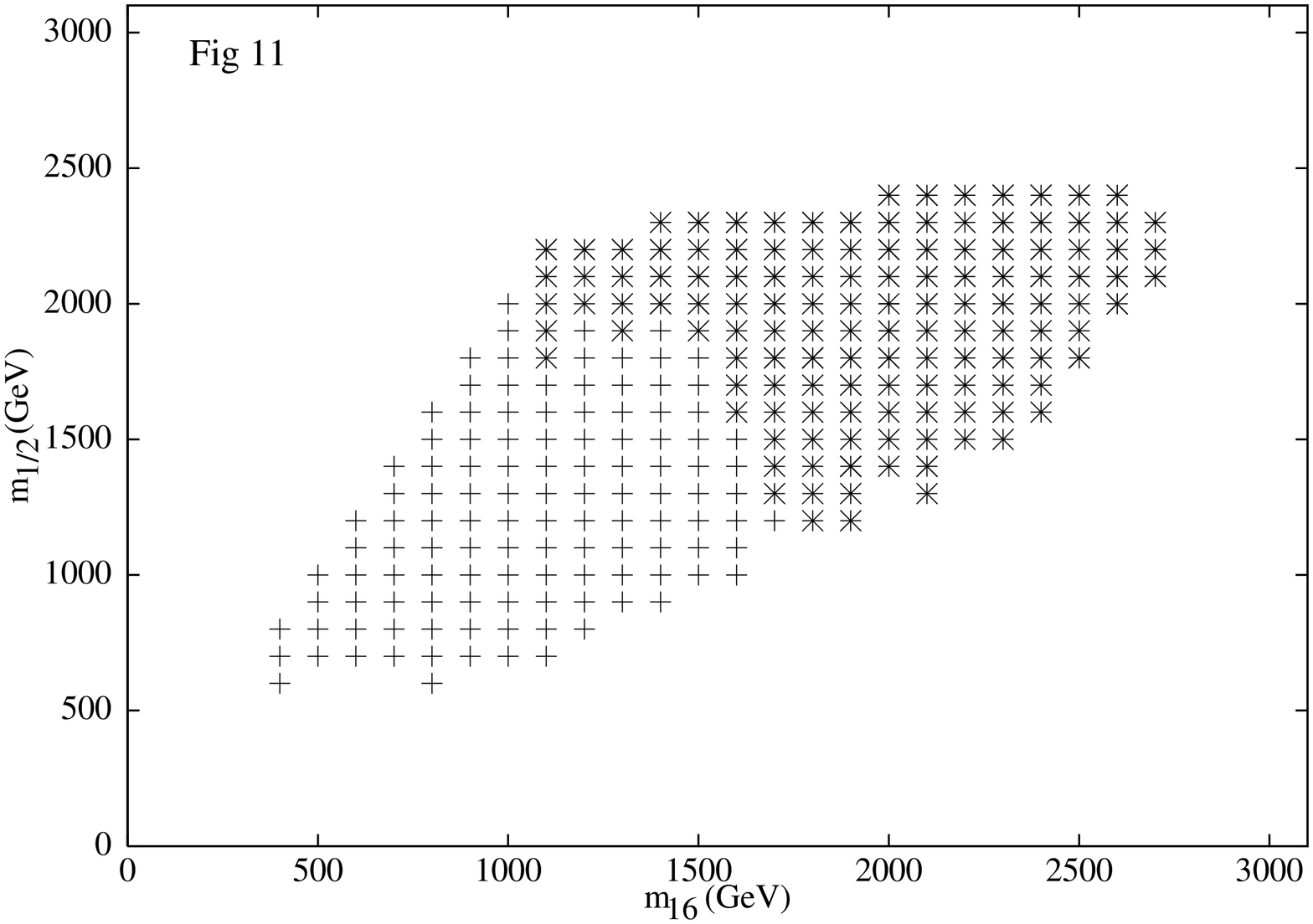,width=12cm,angle=270}
}
\caption{\sl{The allowed parameter space with $t-b-\tau$ unification. Yukawa
             unification to 10\% allows all points, while the asterisks are
             ruled out by UFB-1. We set $m_{10}=m_{16}$ and $A_0=0.3m_{16}$. 
            }}
    \label{fig:figure11}
\vspace*{5mm}
\centerline{
\psfig{file=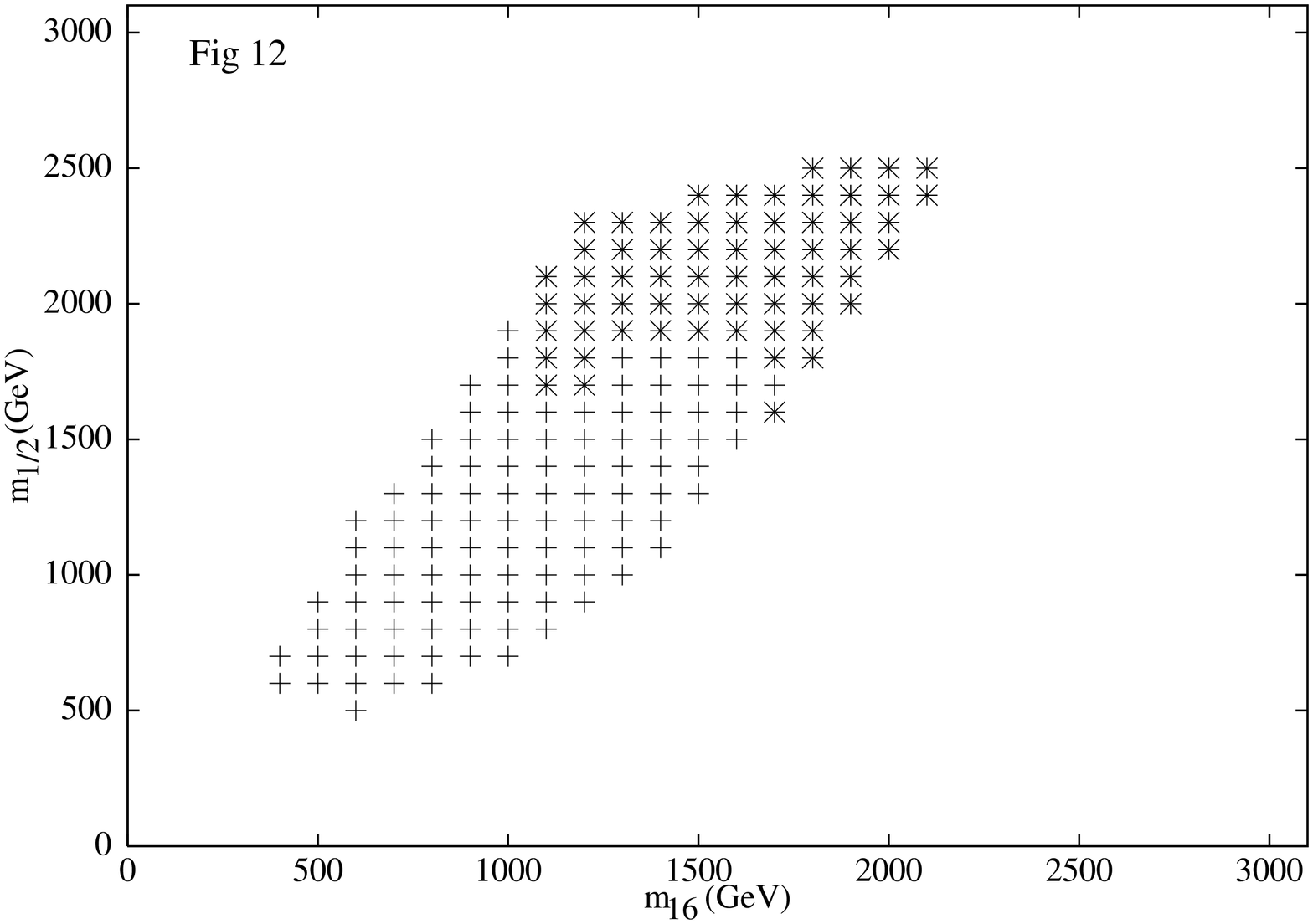,width=12cm,angle=270}
}
\caption{\sl{The same as Fig. 11, with $m_{10}=1.2m_{16}$.
           }}
     \label{fig:figure12}
\end{figure}

We  begin our discussion for the allowed parameter space in the $\MHF$-$\MSX$
plane for\linebreak 
\mbox{$A_0=-2m_{16}$} (see fig.\ 6) in the conventional scenario. 
At each point $\mu$ and $\tan\beta$ have been fixed by the radiative 
electroweak breaking condition and  $b-\tau$ Yukawa unification 
(at an accuracy of 5\%) respectively.  As expected from the discussions 
of the last section, the UFB-1 condition severely restricts the APS
for relatively large $\MHF$ and $\MSX$.  For smaller values of these
parameters, the UFB-3 condition takes over; it is interesting to note 
that for relatively small $\MHF$ and $\MSX$, relevant for SUSY searches
at the LHC, this condition rules out a
small but interesting region of the parameter space. As a result for 
each $\MHF$ there is an upper limit on $\MSX$ and vice versa. 
Thus for $\MSX$ = 500 (700, 900) GeV we find the gluino mass $\MGL$ to be 
definitely less
than 749 (1189, 1820) GeV respectively. It may be recalled that in 
the conventional scenario there is already a lower limit of 
approximately 300 GeV on $\MGL$ from the direct searches at the Tevatron 
\cite{tev}. On the otherhand, for $\MHF$ = 200 (400, 800) GeV both upper
and lower bounds on $\MSX$ emerge, and we get 
590 GeV (1010, 1775) $<\MSQ<$ 1170 GeV (1690, 2200) where $\MSQ$ is the
average squark mass.
Once SUSY signals are seen at the LHC this highly predictive model can 
be tested.

It may be argued that the accuracy to which Yukawa unification holds
is worse than 5\% due to the uncertainties discussed in the introduction. 
Relaxing the accuracy to 10\% the region of the parameter space
allowed by Yukawa unification alone expands. On imposing various stability 
conditions, we find that the UFB-1 constraints become somewhat weaker. 
However, the additional points allowed, especially the ones for
low $\MSX$, are disallowed by the UFB-3 condition
which becomes stronger in this case.  As a result the upper bounds on $\MHF$ 
for relatively small values of  $\MSX$ presented in the last paragraph remain 
more or less unaltered. 

For smaller negative values of $A_0 = -\MSX$ the UFB constraints 
become less effective as may be seen from fig.\ 7. However, the APS is 
already quite restricted due to the requirement of Yukawa unification alone
(this is the complementarity that we have talked about in the introduction).
Although the bulk of the restricted APS can be probed at the LHC, a 
significant region remains inaccessible to it. 

As we keep on increasing $A_0$ (in an algebraic sense) the UFB conditions 
start losing their effectiveness. For $A_0=0$ none of  
these conditions have any further usefulness in constraining the APS;
see fig.\ 8.
However, the stranglehold of Yukawa unification on the APS 
suffices by itself to predict a restrictive mass spectrum. The $m_{16}$-$m_
{1/2}$ plot is bounded from both below and above, and a significant part
of this APS can be probed at the LHC.

For $A_0>0$ the UFB conditions become ineffective. Yukawa unification alone 
yields a loosely restricted APS but most of it lies beyond the kinematic 
reach of the LHC.

\vspace*{2mm}

We next focus our attention on the nonuniversal scenario $\MTN \neq \MSX$.
For a given $A_0$, the parameter space allowed by $b-\tau$ Yukawa 
unification expands considerably from the conventional scenario
for $\MTN < \MSX$. This is illustrated for $A_0 = -2 \MSX$ 
in fig.\ 9 which should be compared with fig.\ 6.
In this case Yukawa unification is achieved for relatively
low $\tan\beta$, which in turn makes $\MHU$ less negative and hence the UFB-1 
constraint weaker to some extent. However, many of the new points so allowed 
are eaten up by the UFB-3 condition.  
As a result, there is an upper bound on the allowed values of $\MHF$ for the
range of $\MSX$ studied by us $(\MSX<3$ TeV). Moreover, the gluino is most 
likely to be observed at the LHC for this {\em entire} range.  Also the
theoretical lower bound on $\MSX$ gets stronger. However, the UFB conditions 
become ineffective as the magnitude of $A_0$ is reduced keeping its sign
negative. At $A_0=0$ hardly any point is ruled out by these UFB conditions.
This trend is similar to what we obtained for the conventional scenario.

On the otherhand, for $\MTN>\MSX$  the APS due to Yukawa
unification alone is reduced quite a bit.
This is illustrated in fig.\  10 with $m_{10}=1.2m_{16}$, 
which should be compared with figures 6
and 9. For relatively large $\MSX$, UFB-1 is a strong constraint as 
before, while some portion in the low $\MSX$ region may be ruled out 
by the UFB-3 condition. We see that the lower bound on $\MSX$ is significantly
weaker than that in the previous case and $\MSX$ as low as 300 GeV is allowed. 
The upper bound on $\MHF$ is also weakened considerably. Yet an observable 
gluino is predicted over most part of the APS. 

\vspace*{2mm}

We now consider the scenario with $t-b-\tau$ Yukawa unification (within an 
accuracy of 10\%) in the conventional scenario.
The UFB-1 condition completely rules out the APS allowed  
by the unification criterion alone for $A_0\leq 0$. (UFB-2 and UFB-3
conditions do not play any major role in constraining the APS.)
For $A_0>0$, the APS (allowed by Yukawa unification) expands gradually;
though a portion of it is ruled out by the UFB-1 constraint, a significant
amount still remains allowed, and a sizable fraction of it is accessible
at the LHC. The UFB-1 condition gets weaker as we go to larger values of
$A_0$. In fig.\ 11, we show the allowed region for $A_0=0.3\MSX$ and 
$\MHF=\MSX$; in fig.\ 12, we introduce nonuniversality by setting 
$m_{10}=1.2m_{16}$. Note that in the latter case the APS allowed by 
Yukawa unification alone is somewhat smaller than that in the conventional
MSUGRA scenario.

Lastly, if the accuracy of the Yukawa unification is reduced to some extent 
(say, to 20\%) the APS allowed by the unification criterion alone is 
significantly enhanced. However, the UFB-1 constraint rules out a large 
amount of this space, and only a small portion survives for negative
$A_0$.

To summarize, the APS for large negative $A_0$ is so restricted by the 
UFB conditions that one should be able, with a bit of luck,
to test the Yukawa unification models that we have discussed at the LHC
by checking the squark and gluino masses. This restriction weakens if one
goes to algebraically larger values of $A_0$. The quantitative nature 
obviously depends on the model chosen.
 
\section{Conclusions}

We have analyzed the consequences of both $b-\tau$ and $t-b-\tau$ Yukawa 
unifications in conjunction with the UFB conditions in the MSUGRA scenario.
In the forrmer case, for $A_0 < 0$, these two constraints nicely 
complement each other in
restricting the APS; when one is weak, the other is sufficiently strong
(see figures 6 and 7). For $A_0\approx 0$ the UFB constraints are 
rather weak. 
However, the requirement of Yukawa unification at an accuracy less than $5\%$  
by itself squeezes the APS sufficiently. As a result, both $m_{1/2}$ and 
$m_{16}$ are bounded (see fig.\ 8 for details) from below as well as 
above. Bulk of this restricted APS is within the striking range of the LHC.
For large positive values of $A_0$, both the UFB conditions and the Yukawa
unification constraint weaken and a large region of the parameter space
accessible at the LHC is permitted.

The most restrictive model that we have studied is the one 
with $A_0=-2m_{16}$. Here, mainly due to the UFB-3 constraint, one obtains
$m_{\tilde g}\lapp 2$ TeV for $m_0\lapp 1$ TeV. Such gluinos are obviously
within the reach of the LHC. For $A_0>0$, the UFB constraints lose their
effectiveness and the loosely restricted APS is rather large. 

If the accuracy of $b-\tau$ unification is relaxed, the APS tends to increase
as expected. However, the upper bound as mentioned above, {\em viz.}, 
$m_{\tilde g}\lapp 2$ TeV for $m_0\lapp 1$ TeV more or less holds for large
negative $A_0$, thanks to the UFB-3 condition.

The requirement of Yukawa unification is less effective in the nonuniversal 
scenario with $m_{10}<m_{16}$. Nevertheless the model on the whole is quite
restrictive due to the UFB constraints. This is illustrated in fig.\ 9  
for $m_{10}=0.6m_{16}$ and $A_0=-2m_{16}$. Here a gluino 
observable at the LHC is almost definitely
predicted for $m_{16}\lapp 3$ TeV. On the 
otherhand, for $m_{10}>m_{16}$, Yukawa unification by itself strongly 
constrains the APS (see fig.\ 10) and the UFB constraints play a 
subdominant role. Again $m_{\tilde g}$ is predicted to be observable at the
LHC over most of the APS.

The masses of the third generation of sfermions are expected to be considerably
lower than that of the first two generations for large values of $\tan\beta$.
In fig.\ 5 we display in the $b-\tau$ unification scheme, alongwith 
the UFB conditions,  the masses of 
the lighter stop ($\tilde t_1$), sbottom ($\tilde b_1$) and stau ($\tilde
\tau_1$) mass eigenstates as functions of $A_0$. They are indeed found to be
considerably lighter than the sparticles belonging to the first two 
generations; in fact, the lighter stop could very well be the second lightest
SUSY particle. Thus in spite of the restrictions imposed by the UFB conditions
and Yukawa unification, light third generation sfermions can be accomodated.
In particular, the possibility that the lighter stop is the NLSP is open for
large negative $A_0$. 

The $t-b-\tau$ Yukawa unification models, with a unification accuracy of 10\%,
are definitely ruled out for $A_0\leq 0$ in the conventional scenario. For
positive values of $A_0$, the UFB-1 condition is less severe, and a 
portion of the parameter space remains allowed, of which a sizable fraction
should be accessible at the LHC. If we relax the accuracy for unification,
the APS increases, most of which could be ruled out by the UFB-1 
condition.

\vspace*{5mm}
 
\centerline{\bf Acknowledgements}

AD thanks Prof.\ E. Reya for hospitalities at the University of Dortmund.
He further thanks both E. Reya and M. Gl\"uck for many discussions on the
dangerous directions of the scalar potential in supersymmetric theories,
and A. Stephan for helps in computation.
His work was supported by DST, India (Project No.\ SP/S2/k01/97)
and BRNS, India (Project No.\ 37/4/97 - R \& D II/474).
AS acknowledges CSIR, India, for a research fellowship.

\end{document}